# Ferromagnetism in Magic-angle Twisted Bilayer Graphene: A Monte Carlo Study

Hanxiang Zhang (张瀚翔)[1,2*], Yuanxiang Gao (高远翔)[3], Z.J. Ding (丁泽军)[2,3#]

[1]*School of the Gifted Young, University of Science and Technology of China, Hefei, Anhui 230026, China*

[2]*Department of Physics, University of Science and Technology of China, Hefei, Anhui 230026, China*

[3]*Hefei National Research Center for Physical Sciences at Microscale, University of Science and Technology of China, Hefei, Anhui 230026, China*

*zhanghx@mail.ustc.edu.cn

#zjding@ustc.edu.cn

**Abstract**: Ferromagnetism emerges when the Moiré superlattice formed by stacking two graphene monolayers in a magic twist angle are filled with integer number electrons. This work investigates the ferromagnetism based on the Ising models for a triangular lattice at one-quarter filling, a square lattice at half filling and a Kagomé lattice at three-quarters filling of electrons. The temperature dependent heat capacity, magnetic susceptibility, energy and magnetization curves are calculated at zero magnetic field with a Monte Carlo method, leading to derive the phase transition temperatures, $T_c$ =0.76, 1.33 and 4.75 K, respectively. Magnetization curves at finite magnetic field show strong hysteresis at temperature below 0.5 K for all the fillings considered, indicating the ferromagnetism of the system; the results are in agreement with experimental observations.

# I. INTRODUCTION

When two graphene monolayers are stacked together with misalignment of a small magic angle (~ 1.1°), strong correlation between electrons emerge due to the flat energy band near the Fermi level.[1-6] This gives rise to a plethora of exotic physical phenomena, such as, correlated insulator phase, superconductivity and ferromagnetism.[7-11] Recent experiments have shown that when the Moiré superlattice is filled with one electron (i.e., one-quarter filling) ferromagnetism emerges at temperature below 1 K,[10] and, near three-quarters filling the system exhibits quantum Hall ferromagnetism, displaying several jumps and plateaus on the hysteresis loops.[11] There also exist some theoretical works for prediction of the ferromagnetic phases in magic angle twisted bilayer graphene (MATBG) when each Moiré unit cell is half-filled.[12-13]

Apart from magneto-transport experiments, ferromagnetism in MATBG can also be explored by means of numerical simulation based on the Ising model. Because the Fermi velocity in the system vanishes to zero at the magic angle,[1] thus filling electrons can be regarded as localized instantly, verifying the validity of the Ising model approach. And since it is a two-dimensional crystal, one can obtain the exact solution of the Ising model in MATBG, making it possible to provide more insights into the mechanisms for ferromagnetism in MATBG and other strongly correlated systems. Previous works have explored ferromagnetism in monolayer graphene and bilayer graphene using Monte Carlo simulation,[14,15] demonstrating the effectiveness of Monte Carlo simulation to such a MATBG system.

In this paper we present a Monte Carlo study on the Ising modelling of MATBG at integer filling number of electrons, $\nu = 1, 2, 3$. We have at first calculated heat capacity, magnetic susceptibility, energy and magnetization as functions of temperature in the absence of external magnetic field to derive phase transition temperature. By sweeping the applied magnetic field at fixed temperature, we observed hysteresis in the phase diagrams of the system.

# II. THEORETICAL MODEL

Here we investigate ferromagnetism in MATBG by using the Ising model including only the nearest neighbour interaction. The Hamiltonian of the system is written as,

$$H = -J\Sigma_{<i,j>}\sigma_i\sigma_j - \mu_B h \Sigma_i \sigma_i \quad (1)$$

where $\sigma_i$ (= $\pm 1$) stands for the spin of a filling electron, $J$ is the interaction energy between the nearest spins and $h$ is the applied external magnetic field. Fig. 1(a) shows a typical schematic of MATBG, wherein the periodic Moiré pattern is obvious.

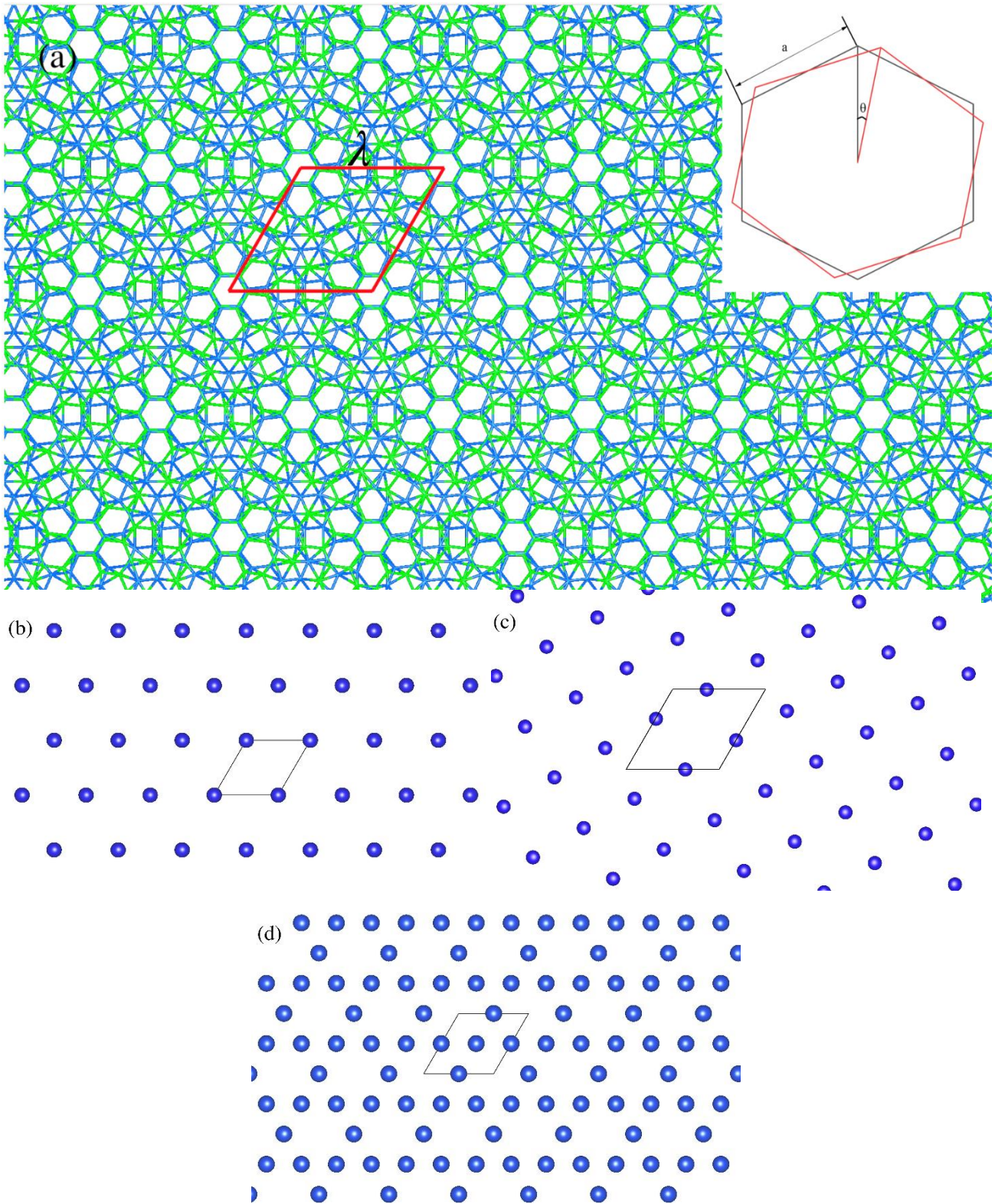

FIG. 1 (a) Moiré superlattice is formed by stacking two monolayers of graphene with a small twist angle. A Moire unit cell is labeled in red solid line. $\lambda = a/2\sin(\theta/2)$ is the moire wavelength where $\theta$ is the twist angle, and $a$ is the lattice constant of graphene. Inset shows the system on a smaller scale. Electron configuration in one Moire unit cell at (b) ν=1 (one-quarter filling) which forms a triangular lattice; (c) ν=2 (half-filling) which forms a square lattice and (d) ν=3 (three-quarters filling) which forms a Kagomé lattice.

At each $\nu$, Moiré unit cell is filled with different number of electrons, and we considered the system taking the configuration with the highest symmetry, as shown in Figs 1(b)-1(d). Clearly, the system forms a triangular lattice, a square lattice and a Kagomé lattice at $\nu = 1, 2$ and $3$, respectively.

In our approach we estimated the interaction energy as $J = 18\ \mu eV$ at $\nu = 1$ from experimental data.[10] According to this value at one-quarter filling, we took $J = 44.8\ \mu eV$ at $\nu = 2$ and $J = 144\ \mu eV$ at $\nu = 3$ since the interaction energy between two magnets is inversely proportional to the distance cubed.

Using the Metropolis sampling algorithm in a Markov chain Monte Carlo simulation, we have considered $N = 10^6$ Moiré unit cells. In a Monte Carlo step, we randomly choose an electron for flipping its spin according to the transition probability, $\exp[-\beta(E_2 - E_1)]$, where $\beta = 1/(k_B T)$, $E_1$ and $E_2$ are the system energies before and after the transition, $T$ the temperature and $k_B$ the Boltzmann constant. At each temperature, 60000 Monte Carlo steps were carried out and the first 20000 steps were discarded to ensure the system is at thermal equilibrium.

At $h = 0$, we investigated the second order phase transition of the system. For each temperature the system was characterized by: energy per spin, $E = \frac{\Sigma_\alpha <U>_\alpha}{2N}$, where $<U>_\alpha$ means magnetization per spin in configuration $\alpha$, the number 2 on the denominator is needed to avoid counting one single spin twice. heat capacity per spin, $C = \frac{(\Delta E)^2}{k_B T^2}$, where $(\Delta E)^2$ is the variance of energy per spin $E$. magnetization per spin, $M = \frac{\mu_B \Sigma_\alpha <\sigma>_\alpha}{N}$, where $<\sigma>_\alpha$ means magnetization per spin in configuration $\alpha$. and magnetic susceptibility per spin, $\chi = \frac{(\Delta M)^2}{k_B T}$.

When an external magnetic field perpendicular to the lattice plane was applied, we fixed the temperature and investigated only the magnetization per spin in the system because the hysteresis loop in magnetic field dependence of magnetization is the key signature of ferromagnetism.

## III. RESULTS AND DISCUSSIONS

First, we examined the behavior of our system at zero external field. In Fig. 2(a) the heat capacity and magnetic susceptibility against temperature at one-quarter filling are plotted. Each curve has a sharp peak at transition temperature $T_c = 0.76 K$, then quickly decreases to zero on both sides. Fig. 2(b) shows the magnetization per spin against temperature. From Fig. 2(b) we observe that the system experiences a phase transition, from ferromagnetic state corresponding to the plateau at temperatures near zero, to paramagnetic state corresponding to the nearly 0 magnetization above $T_c$ in Fig. 2(b). Between the two states the magnetization curve drops quickly, confirming this phase transition is a second order phase transition. And in Fig. 2(b) energy per spin is plotted, showing the system evolving to equilibrium when temperature increases from near 0 to above $T_c$.

Heat capacity and magnetic susceptibility against temperature at half filling are plotted in Fig. 3(a). Similar to their counterparts at one-quarter filling, they exhibit a peak on each diagram, but at a different transition temperature $T_c = 1.33K$. This result is natural since at $\nu = 1$ and $\nu = 2$ the system takes different electron configurations. The transition temperature is higher due to the enhanced spin interaction. In Fig. 3(b) the temperature dependence of magnetization per spin and energy per spin against is plotted, showing the system evolving from an ordered state near zero temperature to a thermodynamic equilibrium state at high temperatures.

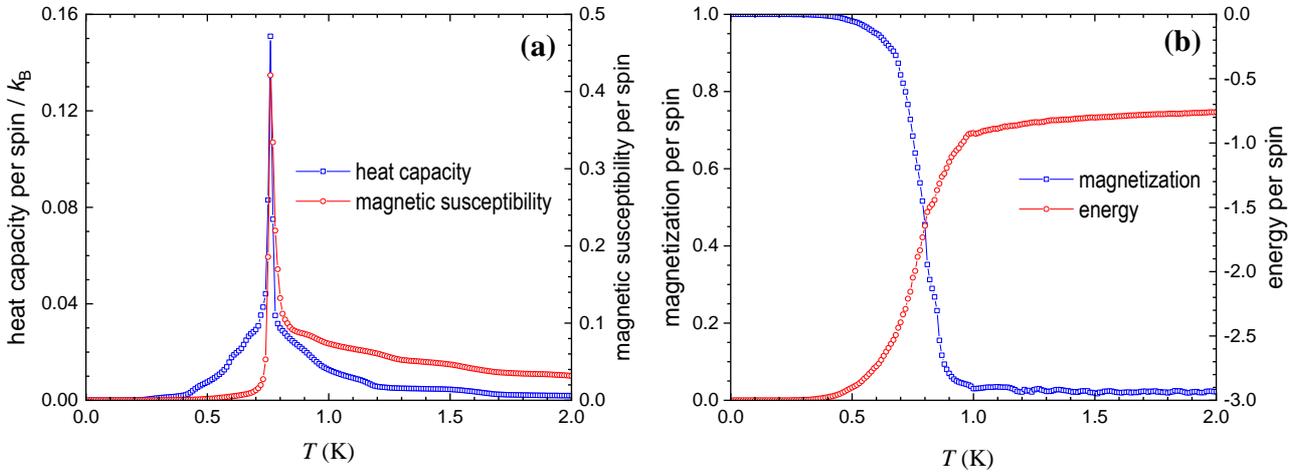

FIG. 2 Temperature dependence of system's (a) heat capacity and magnetic susceptibility, from the peak in both diagram we obtained transition temperature of the system $T_c = 0.76K$. (b) magnetization per spin and energy per spin in terms of J against temperature T at $\nu = 1$.

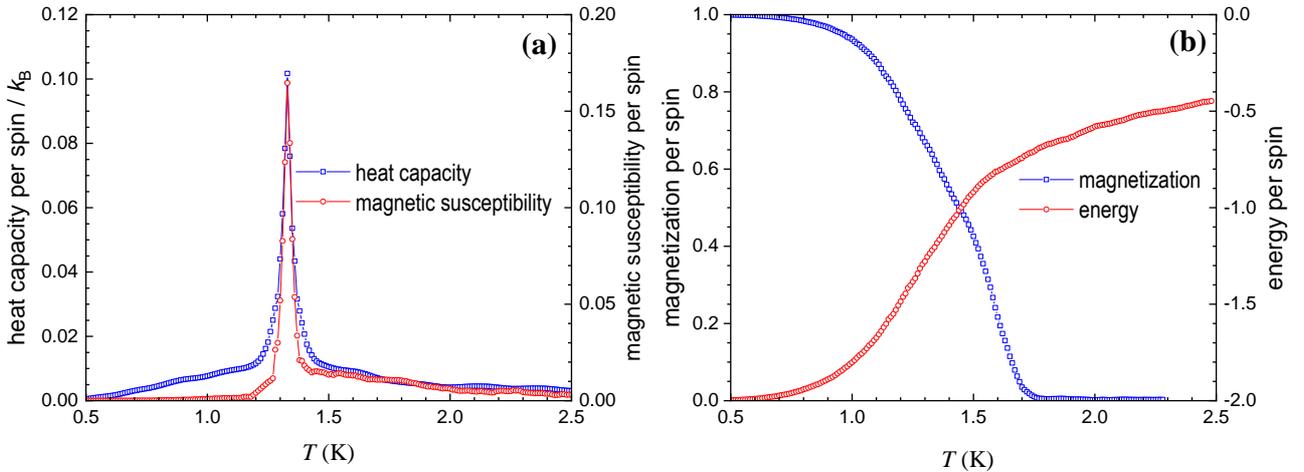

FIG. 3 Temperature dependence of system's (a) heat capacity and magnetic susceptibility, from the peak in both diagrams we obtained transition temperature of the system $T_c = 1.33K$. (b) magnetization per spin and energy per spin in terms of J against temperature T at $\nu = 2$.

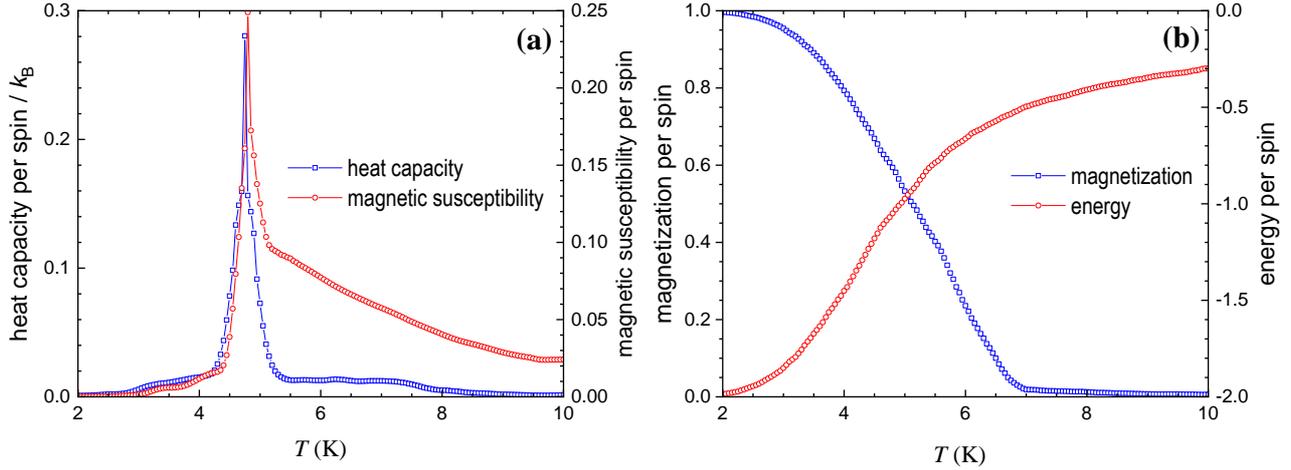

FIG. 4 Temperature dependence of system's (a) heat capacity and magnetic susceptibility, from the peak in both diagrams we obtained transition temperature of the system $T_c = 4.75K$. (b) magnetization per spin and energy per spin in terms of J against temperature T at $\nu = 3$.

The corresponding temperature dependence of the system at three-quarters filling are plotted in Fig. 4. Under this scenario, we extracted $T_c = 4.75K$ from curves in Fig. 4(a) which is much higher than previous cases due to the enhanced spin interaction at $\nu = 3$, which is 8 times larger than the one-quarter filling case. It is also noted that the thermal fluctuation is more obvious evinced by wide peaks in Fig. 4(a). We contributed this to the strong spin interaction in the system, making it hard for a large number of spins to flip as a whole. In Fig. 4(b) the temperature dependence of magnetization per spin and energy per spin against temperature is plotted, showing the system evolving from a ferromagnetic phase near zero temperature to a paramagnetic state at high temperatures above $T_c$.

We then investigated the magnetization of the system when a finite external field is applied. Figure 5(a) shows the magnetic field dependence of magnetization at one-quarter filling at different temperatures, T=0.1K, 0.5K 0.76K ($T_c$) and 1.2K. The hysteresis loops are conspicuous below 0.5K, indicating the ferromagnetic phases of the system. At 0.76K ($T_c$), the hysteresis loop is distorted and at 1.2K it can not be seen and thus indicating the system experience a phase transition from ferromagnetic phase to paramagnetic phase. Our simulation results are in agreement with experimental results.[10]

In Fig. 5(b), the magnetic field dependence of magnetization at half filling at different temperatures, T=0.1K, 0.5K 1.33K ($T_c$) and 2K, are plotted. The hysteresis loops exhibit similar behaviors to ones in Fig. 5. But the critical field at which the spins flip as a whole is bigger. This result is understandable because the spin interaction is stronger in this case. At 1.33K ($T_c$), the hysteresis loop is distorted and at 2K it can not be seen and thus indicating the system experience a phase transition from ferromagnetic phase to paramagnetic phase. Our simulation results agree well with theoretical predictions.[14]

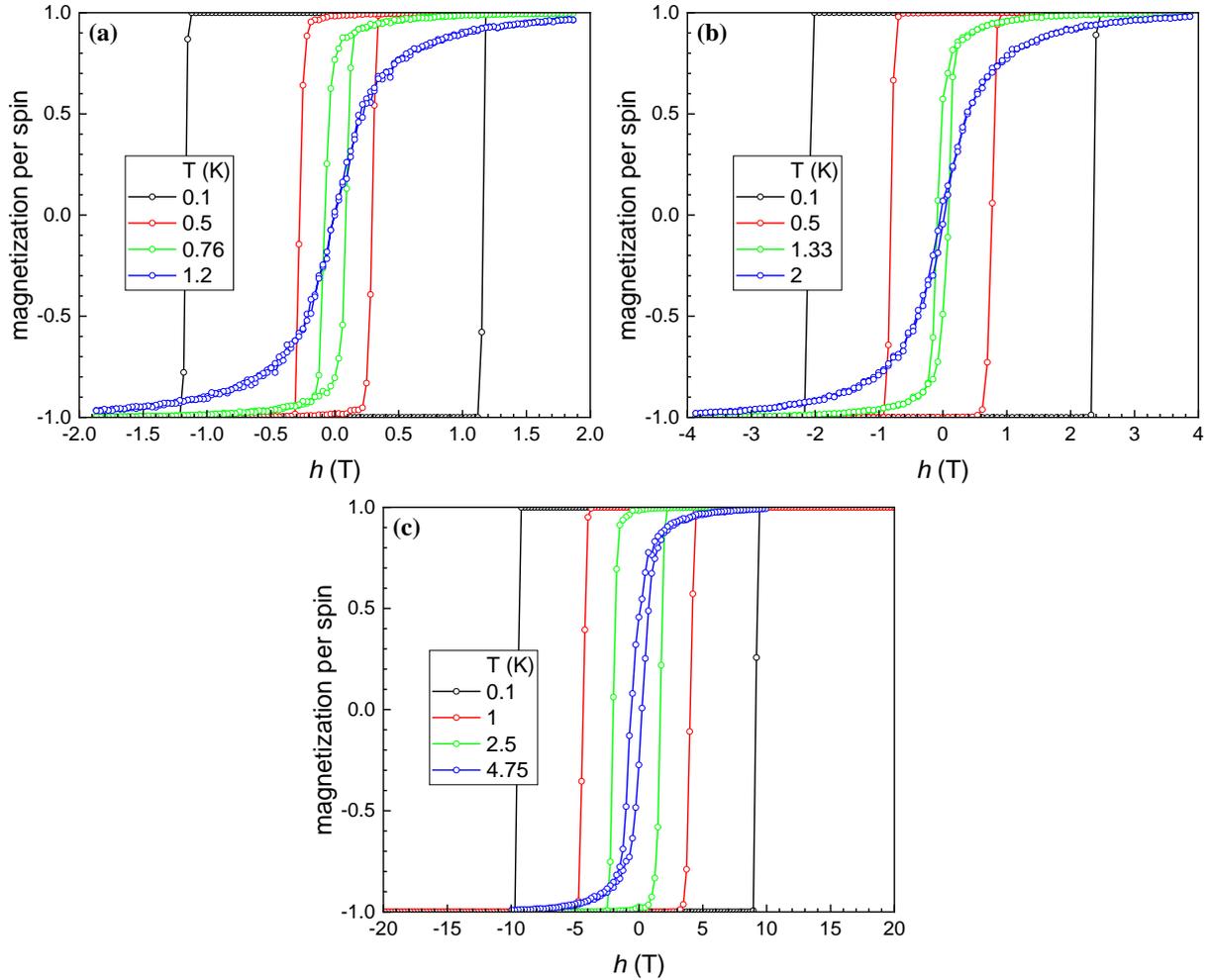

Figure 5 (a) ($\nu = 1$) Magnetization curves when increasing and decreasing external magnetic field is applied under T=0.1K, T=0.5K, T=0.76K which is the transition temperature, and T=1.2K.
(b) ($\nu = 2$) Magnetization curves when increasing and decreasing external magnetic field is applied under T=0.1K, T=0.5K, T=1.33K which is the transition temperature, and T=2K.
(c) ($\nu = 3$) Magnetization curves when increasing and decreasing external magnetic field is applied under T=0.1K, T=1K, T=2.5K and T=4.75K which is the transition temperature.

In Fig. 5(c), the magnetic field dependence of magnetization at three-quarters filling under different temperatures are obtained. The hysteresis loop is conspicuous below 2.5K. It is obvious that under same temperature, in order to flip over the spins of the whole system, a larger magnetic field is required than the one-filling case due to the stronger interaction between electrons at $\nu = 3$.

It is noted that we do not observe anomalous quantum Hall effect in our hysteresis loops. This is mainly because the underlying symmetry-broken correlated state in which AQHE was observed[11] was induced by the coupling between the graphene device and its encapsulating substrate,[13] which was hexagonal boron nitride (hBN) in this case. In our model we only considered the MATBG itself, precluding the coupling between MATBG and hBN substrate, leading to the absence of AQHE.

## IV. CONCLUSIONS

In conclusion, we have established the Ising model of MATBG at integer filling number $\nu = 1,2,3$. Our model showed that at one-quarter filling the electron configuration takes the triangular lattice, at half filling the electron configuration takes the square lattice, and at three-quarters filling the electron configuration takes the Kagomé lattice. Employing Monte Carlo method and Metropolis sampling algorithm, we showed that transition temperature is $0.76K, 1.33K, 4.75K$ at $\nu = 1,2,3$ respectively. Our simulation results of magnetization of the system under finite external magnetic field exhibited strong hysteresis below $0.5K$, indicating ferromagnetism in the system, and a phase transition from ferromagnetic to paramagnetic when temperature is increased, which agrees well with experimental data. However, due to the strong fluctuation in the system near transition temperature resulting in distortion of curves in our diagrams, more simulations are required to offset its effect. It should also be noted that the Ising model is the simplest description of ferromagnetism and it often overestimates the transition temperature of the system, thus a study on a more realistic model, the Heisenberg model as an example, would provide more insight into the MATBG system and other strongly correlated system.

## ACKNOWLEDGEMENTS

This work was supported by the Chinese Education Ministry through the "111" Project2.0 (BP0719016).